\documentclass[%
 reprint,
longbibliography,
 amsmath,amssymb,
 aps,
pra,
]{revtex4-1}
\usepackage{graphicx}
\usepackage{dcolumn}
\usepackage{bm}
\newcommand{\ud}{\mathrm{d}}

\begin{document}

\preprint{APS/123-QED}

\title{Impurity coupled to an artificial magnetic field in a Fermi gas in a ring trap}

\author{F. Nur \"{U}nal}
\email{fatmanur@bilkent.edu.tr}%
\author{B. Het\'{e}nyi}%
\author{M. \"{O}. Oktel}
\affiliation{Department of Physics, Bilkent University, Ankara, Turkey}




\date{\today}

\begin{abstract}
The dynamics of a single impurity interacting with a many particle background is one of the central problems of condensed matter physics. Recent progress in ultracold atom experiments makes it possible to control this dynamics by coupling an artificial gauge field specifically to the impurity. In this paper, we consider a narrow toroidal trap in which a Fermi gas is interacting with a single atom. We show that an external magnetic field coupled to the impurity is a versatile tool to probe the impurity dynamics. Using Bethe Ansatz (BA) we calculate the eigenstates and corresponding energies exactly as a function of the flux through the trap. Adiabatic change of flux connects the ground state to excited states due to flux quantization. For repulsive interactions, the impurity disturbs the Fermi sea by dragging the fermions whose momentum matches the flux. This drag transfers momentum from the impurity to the background and increases the effective mass. The effective mass saturates to the total mass of the system for infinitely repulsive interactions. For attractive interactions, the drag again increases the effective mass which quickly saturates to twice the mass of a single particle as a dimer of the impurity and one fermion is formed. For excited states with momentum comparable to number of particles, effective mass shows a resonant behavior. We argue that standard tools in cold atom experiments can be used to test these predictions.
\begin{description}
\item[PACS numbers]
\end{description}
\end{abstract}

\pacs{Valid PACS appear here}
\maketitle


\section{Introduction}
Ultracold atom systems are effectively used as a test bed for condensed matter models. They are preferred because of the high degree of control in experiments such as tunable interactions, impurities and direct measurements by optical techniques. Certain theoretical models of condensed matter such as resonant interactions \cite{resonance} or bosonic Mott transition \cite{mott} have been realized for the first time using cold atoms. Many models of one dimensional systems have been realized using two dimensional optical lattices to form narrow tubes \cite{1DTubeWeiss,1DTubeParades,1DTubeEsslinger}.

One of the powerful theoretical tools to describe one dimensional systems is the Bethe Ansatz (BA). BA solution has been generalized to many integrable models, e.g. systems with multiple components, different statistics or spin \cite{Bethe,LiebLiniger,Yang,McGuire,GuanReview}. This exact solution method has been employed to explain experimental data on a number of instances \cite{1DTubeParades,MeanfieldBreakdown}. However, as BA methods are restricted to one dimension, they have not been used to describe systems where an external artificial gauge field is present.

In one dimension, such an external magnetic field can be disregarded by using a gauge transformation, unless the one dimensional system closes onto itself. Thus, if the particles are confined to a ring as opposed to a tube, the artificial magnetic field will significantly effect the physics. Such rings, in the form of toroidal traps, have been realized experimentally \cite{StamperKurnCircularWaveguide,StamperKurn2015,RingCJFoot,RingSchmiedmayer,RingChapman,RingPrentiss,RingRiis,RingCampbell}. Although none of these experiments have included an artificial gauge field so far.

In this work, we consider such a toroidal trap containing non-interacting fermions and describe the behavior of a single charged impurity interacting with background atoms. We argue that an artificial magnetic field coupling to the impurity is an efficient way to probe the polaron state forming due to the interactions. Artificial magnetic fields are created by coupling light to the internal states of the atoms \cite{spielman,KetterleArtificial,BlochHofstadter}. Hence, they are highly specific to the internal state making it possible to create effective magnetic fields coupling only to one type of atom.

The charged particle is expected to drag the uncharged fermions along with itself around the ring. Because of the interactions between the impurity and the background atoms a collective excitation usually called a polaron is formed \cite{Polaron}. This excitation will couple to the external magnetic field with the charge of the impurity particle, however, its mass will critically depend on the interaction strength. The amount of angular momentum carried by the impurity and the uncharged fermions also depend on the total external flux through the ring. By changing the artificial magnetic field strength, it is possible to access excited states of the system adiabatically. We show that an artificial magnetic field coupling specifically to the impurity would be a very effective tool to probe polaron physics.

We describe this system exactly using a Bethe Ansatz (BA) solution for contact interactions which are justified for cold atoms as the dominant scattering is s-wave. For strongly attractive interactions, the impurity forms a bound state with one of the background fermions and the physics reduces to the motion of a dimer with twice the mass of the particle. In the other limit of infinitely repulsive interaction, effective mass saturates to total particle number. We calculate the energy and momentum distributions, total transferred momentum and the effective mass for all interaction strengths. We believe these results can be experimentally checked with state of the art toroidal traps and techniques for artificial gauge field generation.

The paper is organized as follows. In the next section, we define the model, introduce the notation and review earlier studies. In Section III, we solve the system for two particles and then generalize to any particle number using the BA. Sec. IV contains the analytical solution of the BA equations in certain limits and comparison with numerical solutions. We present our results for several quantities such as energy, angular momentum and effective mass of the charged particle. We give our conclusions along with a brief discussion of possible experiments in Sec. V.
\begin{figure}
\includegraphics[width=0.15\textwidth]{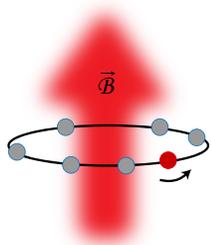}
\caption{ (Color online) A simple illustration of the system. $N-1$ uncharged fermions (light gray) and a single charged impurity (dark gray) are trapped on a ring. The impurity is interacting with the fermions via Delta-function interaction. An artificial magnetic field couples exclusively to the impurity. The dynamics of the system depends on the interaction strength between particles and the total flux through the ring $\beta=\frac{qRA}{\hbar}$.} \label{illusration}
\end{figure}

\section{THE MODEL}
The first quantized Hamiltonian for one charged particle among $N-1$ uncharged fermions under a magnetic field reads
\begin{equation}
{\cal H}=\frac{1}{2m}\Big(\frac{\hbar}{i}\frac{\partial}{\partial x_1}-qA\Big)^2- \frac{\hbar^2}{2m}\sum_{j=2}^N\frac{\partial^2}{\partial x_j^2}+2c\sum_{j=2}^N \delta(x_1-x_j).
\end{equation}
All particles are assumed to be on a ring of radius R, $0\leq x_i \leq 2\pi R$. The position of the charged particle is $x_1$ and $A$ is the vector potential in the symmetric gauge. The Hamiltonian can be made dimensionless by using, $\tilde{x}_j=\frac{x_j}{R}$, $\tilde{E}=E\frac{2mR^2}{\hbar^2}$, $\tilde{c}=c\frac{2mR}{\hbar^2}$ and $\beta=\frac{qRA}{\hbar}$. $\beta$ is the total magnetic flux through the ring in units of flux quantum $q/h$. Dropping the tildes
\begin{equation}
{\cal H}=\Big(-i\frac{\partial}{\partial x_1}-\beta\Big)^2- \sum_{j=2}^N\frac{\partial^2}{\partial x_j^2}+2c\sum_{j=2}^N \delta(x_1-x_j).
\end{equation}
The effect of the magnetic field can be shifted to the boundary conditions by a gauge transformation \cite{ShastrySutherland}. Namely, when the first particle makes a full circle around the ring the wave function gains a phase factor of $e^{i\beta 2\pi}$ where the periodic boundary conditions (PBCs) for the uncharged particles remain unaffected by the gauging process,
\begin{equation}
{\cal H}\rightarrow e^{-i\beta x_1}{\cal H}e^{i\beta x_1}.
\end{equation}

Apart from the twisted BCs, the $\delta$-function interaction can be handled as a two-sided boundary condition (BC) between two different regions of N-particle space corresponding to different permutations of particles. The discontinuity relation at the boundary $x_1=x_j$ (which is obtained by passing to the center of mass and relative coordinates and then integrating the Hamiltonian) is
\begin{equation} \label{discont genel}
(\partial_j-\partial_1)\psi\Big|_{x_1<x_j}-(\partial_j-\partial_1)\psi\Big|_{x_j<x_1}=2c\psi\Big|_{x_j=x_1} ,\quad j\neq1.
\end{equation}

This one dimensional problem of two-component fermions has been studied by using the BA in the previous century. First, the one-spin deviate problem in a Fermi sea is solved by McGuire \cite{McGuire} and Flicker and Lieb \cite{FlickerLieb} solved the two-spin deviate problem. Yang \cite{Yang} elegantly derived the BA equations for the general $M$ down-spins among $N$ up-spins. Twisted BCs have been used throughout the BA literature as a way to probe ground state properties. However, with the possibility of optically inducing artificial magnetic fields, it is important to calculate the properties of the system at finite flux as opposed to infinitesimal values near zero. It is also necessary to consider cases where different components in the system experience different gauge fields.

Our calculation takes both of these constraints into account and allows us to exactly study the dynamics resulting from the dragging effect of the charged particle on the uncharged particles. The resulting polaron physics has attracted great interest in the context of cold atoms over the last few years \cite{Polaron,AnnaMinguzzi}.

\section{THE ANSATZ}
As the interactions are reduced to BCs, the wave function for a given permutation of the particles is a superposition of plane waves. As the collisions of equal mass particles in one dimension conserve magnitudes of the incoming momenta, the interacting problem is integrable. Hence, in a given region only a finite number of plane waves are needed to construct the wave function. To make our notation clear, we first start with the case of one charged particle with one neutral particle.

\subsection{N=2 Particles}
For two particles we have $2!=2$ regions and the wave function in these regions is expressed as follows:
\begin{eqnarray}
\Psi_{12}(x_1,x_2)=(12)_{12}e^{i(k_1x_1+k_2x_2)}+(21)_{12}e^{i(k_2x_1+k_1x_2)},\nonumber\quad\\
\Psi_{21}(x_1,x_2)=(12)_{21}e^{i(k_1x_1+k_2x_2)}+(21)_{21}e^{i(k_2x_1+k_1x_2)},\quad
\end{eqnarray}
where we use parenthesis with a subscript to indicate the coefficients of plane waves. In this notation numbers in the parenthesis indicate the order the wave vectors $k_1,k_2$ are distributed to the coordinates in the exponent and the subscript indices indicate the ordering of the particles on the ring, \textit{i.e.} $\Psi_{12}$ means $x_1<x_2$. At $x_1=x_2$, the wave functions in the two regions should be equal whereas their derivative should obey Eq.\ref{discont genel}. Equating the coefficients of each plane wave on both sides, we obtain:

\textbf{BCs:} at $x_1=x_2$,
\begin{equation}
(12)_{21}+(21)_{21}=(12)_{12}+(21)_{12},
\end{equation}
\begin{equation}
(12)_{21}-(21)_{21}=(12)_{12}\Big(1+\frac{2}{s_{12}}\Big)+(21)_{12}\Big(-1+\frac{2}{s_{12}}\Big), \label{disc.}
\end{equation}
where $s_{12}=i(k_1-k_2)/c$. Combined BCs give
\begin{equation}\label{bla}
\begin{pmatrix}
(12)\\\\(21)
\end{pmatrix}_{21} =
\begin{pmatrix}
1+\frac{1}{s_{12}} & \frac{1}{s_{12}} \\\\
\frac{-1}{s_{12}} & 1-\frac{1}{s_{12}}
\end{pmatrix}
\begin{pmatrix}
(12)\\\\(21)
\end{pmatrix}_{12}.
\end{equation}
Allowed values for $k_1,k_2$ are found by applying the PBCs. PBC for one of the particles gives the BA equation.

\textbf{BCs:} at $2\pi$

as $ x_2:0\rightarrow2\pi,\qquad \Psi_{21}(x_2=0)=\Psi_{12}(x_2=2\pi)$,
\begin{equation}
\qquad(12)_{21}=(12)_{12}e^{ik_22\pi},\qquad (21)_{21}=(21)_{12}e^{ik_12\pi}. \label{pbc2}
\end{equation}

as $ x_1:0\rightarrow2\pi,\qquad \Psi_{12}(x_1=0)=e^{i\beta2\pi}\Psi_{21}(x_1=2\pi)$,
\begin{equation}
\quad(12)_{12}=(12)_{21}e^{i(k_1+\beta)2\pi},\quad (21)_{12}=(21)_{21}e^{i(k_2+\beta)2\pi}. \label{pbc1}
\end{equation}
Combining the two BCs at $2\pi$, we obtain another constraint $k_1+k_2+\beta=n$, for $n\in\mathbb{Z}$. This is a reflection of the total angular momentum conservation in the system. Eqs.\ref{bla} and Eq.\ref{pbc1} have non-trivial solutions only when the determinant below vanishes,
\begin{equation}
\begin{vmatrix}
1+\frac{1}{s_{12}}-e^{-i(k_1+\beta)2\pi} & \frac{1}{s_{12}} \\\\
\frac{-1}{s_{12}} & 1-\frac{1}{s_{12}}-e^{-i(k_2+\beta)2\pi}
\end{vmatrix}=0.
\end{equation}
\begin{figure}
\includegraphics[width=0.47\textwidth]{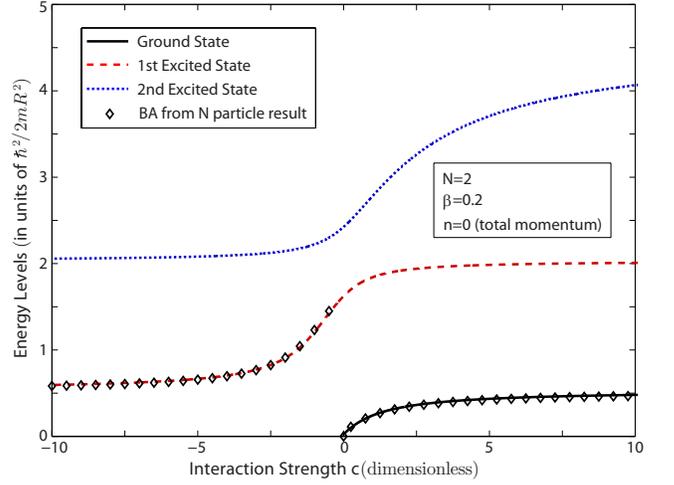}
\caption{ (Color online) Energy of the lowest three states vs. interaction strength for $N=2$ particles, for zero total angular momentum. Only scattering states are displayed. Energy is calculated by three different methods. Lines are from Eq.\ref{Tez Eq.} direct analytical solution without employing BA, which is algebraically same with the two-particle BA calculation (Eq.\ref{BA for N=2}). Diamonds are from the general $N$-particle BA calculation (Eq.\ref{BA eqs}).} \label{EvsC N2 fig}
\end{figure}
Solution of this determinant gives the BA equation,
\begin{equation}
\alpha=\frac{c}{2}\cot\Big(\frac{\pi}{2}(\alpha+n-\beta)\Big)+\frac{c}{2}\cot\Big(\frac{\pi}{2}(\alpha-n+\beta)\Big), \label{BA for N=2}
\end{equation}
where energy is $E=\frac{(n-\beta)^2+\alpha^2}{2}$ for $\alpha=k_2-k_1$. For the two-particle case, this problem can also be solved exactly without using the BA \cite{tez},
\begin{equation}
c=\alpha\Big(\frac{\cos(\pi(n-\beta))}{\sin(\pi\alpha)}-\cot(\pi\alpha)\Big).\label{Tez Eq.}
\end{equation}
These two equations analytically reproduce each other and the numerical results match perfectly (Fig.\ref{EvsC N2 fig}).

Extension of this method to N particles is straightforward if cumbersome.
\begin{figure}
\includegraphics[width=0.47\textwidth]{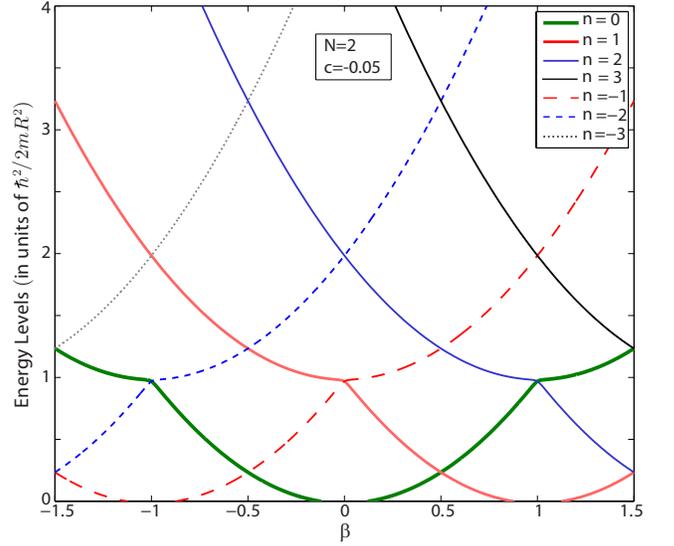}
\caption{(Color online) Ground state energy vs. flux $\beta$ for $N=2$ particles with total angular momentum $n$. Eigenstates for flux $\beta$ with total angular momentum \textit{n} are also eigenstates for flux $\beta+1$ with total angular momentum $n+1$. The system can be analyzed by considering flux values between $-1/2<\beta<1/2$ for all $n$. As the flux is increased by one adiabatically, the system evolves to a higher excited state which has one more unit of total angular momentum. As can be observed, the crossings between different eigenstates is not a problem for adiabatic evolution since only states with different total angular momentum \textit{n} are degenerate.} \label{E vs beta}
\end{figure}

\subsection{N-1 Fermions, One Charged Particle}
The distinguishable charged particle is denoted again by $x_1$ and the wave function is defined in $N!$ regions corresponding to different permutations \cite{McGuire}. In each one of these regions the wave function consists of $N!$ plane waves in its most general form without imposing the antisymmetry between the fermions. As a total we have $N!\times N!$ coefficients:
\begin{widetext}
\begin{eqnarray}
\Psi_{123\ldots}=(123\ldots)_{123\ldots}e^{i(k_1x_1+k_2x_2+k_3x_3+\ldots)}+(213\ldots)_{123\ldots}e^{i(k_2x_1+k_1x_2+k_3x_3+\ldots)}+\ldots\nonumber\\
\Psi_{213\ldots}=(123\ldots)_{213\ldots}e^{i(k_1x_1+k_2x_2+k_3x_3+\ldots)}+(213\ldots)_{213\ldots}e^{i(k_2x_1+k_1x_2+k_3x_3+\ldots)}+\ldots\nonumber\\
\Psi_{132\ldots}=(123\ldots)_{132\ldots}e^{i(k_1x_1+k_2x_2+k_3x_3+\ldots)}+(213\ldots)_{132\ldots}e^{i(k_2x_1+k_1x_2+k_3x_3+\ldots)}+\ldots\nonumber\\
\qquad\vdots\qquad\qquad\qquad\qquad\qquad\qquad\vdots\qquad\qquad\qquad\qquad
\end{eqnarray}
\end{widetext}
where $k_1,k_2,\ldots k_N$ are distinct wavenumbers. BCs at $x_1=x_2$ are not effected by the addition of other fermions at the end of the sequence:
\begin{equation}
\begin{pmatrix}
(123\ldots)\\\\(213\ldots)
\end{pmatrix}_{213\ldots} =
\begin{pmatrix}
1+\frac{1}{s_{12}} & \frac{1}{s_{12}} \\\\
\frac{-1}{s_{12}} & 1-\frac{1}{s_{12}}
\end{pmatrix}
\begin{pmatrix}
(123\ldots)\\\\(213\ldots)
\end{pmatrix}_{123\ldots}.
\end{equation}

BCs at $2\pi$ follow the same logic;

as $ x_2:0\rightarrow2\pi,\; \Psi_{213\ldots N}(x_2=0)=\Psi_{13\ldots N2}(x_2=2\pi)$, yielding
\begin{eqnarray}
\;(123\ldots)_{213\ldots N}=(123\ldots)_{13\ldots N2}e^{ik_22\pi}, \nonumber\\
(213\ldots)_{213\ldots N}=(213\ldots)_{13\ldots N2}e^{ik_12\pi}.
\end{eqnarray}
Number of independent coefficients decreases considerably by requiring antisymmetry upon exchange of fermions. Every coefficient of a plane wave in region $x_1<x_3<x_2<\ldots<x_N$ is identical with the coefficient of the same plane wave in region $x_1<x_2<x_3<\ldots<x_N$. This can be shown by noticing that at $x_2=x_3$ the wave functions must vanish requiring \textit{e.g.} $(123\ldots)_{123\ldots N}=-(132\ldots)_{123\ldots N}$. Fermionic antisymmetry also relates the wave function in separate regions $\Psi_{123\ldots N}=-\Psi_{132\ldots N}$. As a result, the coefficients only depend on the position of the charged particle in the order. We can move indistinguishable fermions through one another at will and the $N!$ regions reduce to $N$ regions.

After this simplification it is easy to combine the BC at a $\delta$-function with the overall PBC.
\begin{eqnarray} \label{Nparticle det.}
\begin{pmatrix}
(123\ldots)\\\\(213\ldots)
\end{pmatrix}_{213\ldots N} &=&
\begin{pmatrix}
1+\frac{1}{s_{12}} & \frac{1}{s_{12}} \\\\
\frac{-1}{s_{12}} & 1-\frac{1}{s_{12}}
\end{pmatrix}
\begin{pmatrix}
(123\ldots)\\\\(213\ldots)
\end{pmatrix}_{123\ldots N} \nonumber\\
&=&\begin{pmatrix}
e^{ik_22\pi}(123\ldots)_{123\ldots N}\\\\e^{ik_12\pi}(213\ldots)_{123\ldots N}
\end{pmatrix}.
\end{eqnarray}
The determinant can only vanish if $k_1$ and $k_2$ satisfy,
\begin{equation}
k_1-\frac{c}{2}\cot{\pi k_1}=k_2-\frac{c}{2}\cot{\pi k_2}.
\end{equation}

The same procedure can be applied to any pair of wavenumbers $k_i,k_j$. Thus, all the wavenumbers must satisfy
\begin{equation*}
k_1-\frac{c}{2}\cot{\pi k_1}=k_2-\frac{c}{2}\cot{\pi k_2}=k_3-\frac{c}{2}\cot{\pi k_3}=\ldots=\lambda,
\end{equation*}
where $\lambda$ is a real constant. This form is equivalent to the usual BA equations \cite{LiebLiniger}.

Hence, the N wavenumbers which define an eigenstate must be chosen as N distinct roots of the equation:
\begin{equation}
k-\lambda=\frac{c}{2}\cot{\pi k}.
\end{equation}
However, there is another constraint. Applying PBCs sequentially on all particles restricts $\lambda$.
\begin{equation*}
\text{As } x_1:0\rightarrow2\pi,\; \Psi_{123\ldots N}(x_1=0)=e^{i\beta2\pi}\Psi_{23\ldots N1}(x_1=2\pi),
\end{equation*}
\begin{equation*}
\;(123)_{123}=(123)_{231}e^{i(\beta+k_1)2\pi},
\end{equation*}
\begin{eqnarray*}
\text{as } x_2:0\rightarrow2\pi, \quad(123)_{231}=(123)_{312}e^{ik_22\pi},\\
\text{as } x_3:0\rightarrow2\pi, \quad(123)_{312}=(123)_{123}e^{ik_32\pi},\\
\vdots\qquad\qquad\qquad\qquad\vdots\qquad\qquad\qquad\qquad
\end{eqnarray*}
In combination:
\begin{equation}
(123\ldots)_{123\ldots N}=e^{i(k_1+k_2+\ldots+k_N+\beta)2\pi}(123\ldots N)_{123\ldots N}
\end{equation}
reflecting angular momentum conservation, sum of all the wavenumbers plus the flux must be integer on a ring.

In short, the BA equation is solved by finding N roots of a simple equation subject to the angular momentum constraint:
\begin{equation} \label{BA eqs}
k-\lambda=\frac{c}{2}\cot{\pi k}, \qquad\sum_j^Nk_j=n-\beta,\quad n\in\mathbb{Z}.
\end{equation}
So far our treatment implicitly assumed repulsive interactions. In which case, all the wavevectors $k_i$ are real. The ansatz can easily be extended to attractive interactions yielding exactly the same equations Eq.\ref{BA eqs} \cite{McGuireAttractive}. However, for negative $c$ two of the roots will be complex, as the $\delta$-potential in one dimension has only a single bound state.
\begin{figure}
\includegraphics[width=0.47\textwidth]{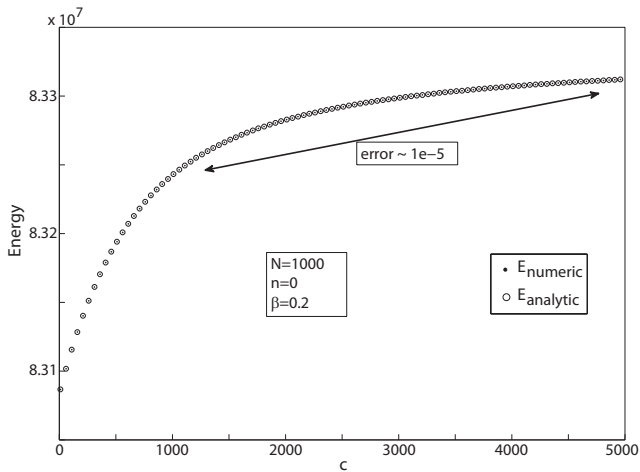}
\caption{ Ground state energy vs. interaction strength for $N=1000$ particles and zero total angular momentum. Numerical solution of BA equation (dots) Eq.\ref{BA eqs} are virtually indistinguishable from analytical solution (circles) $E=$(Fermi energy of $N-1$ fermions)$+\Delta E$. Error between the numerical and analytical solutions are too small to observe even in the regimes where the assumptions for analytical calculation fails. } \label{E vs c}
\end{figure}

\section{SOLUTION OF THE BA EQUATION}
\begin{figure}
\includegraphics[width=0.47\textwidth]{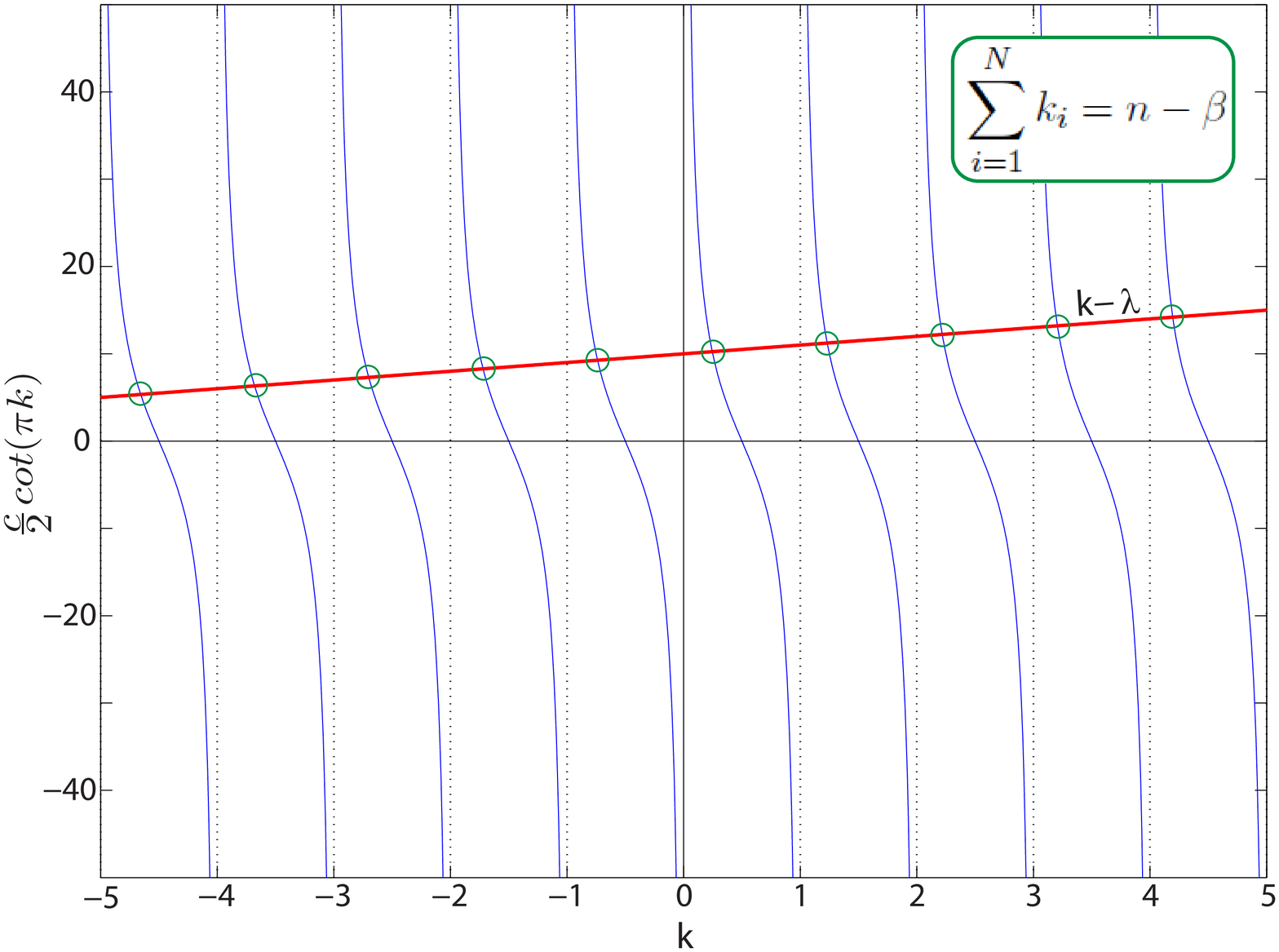}
\caption{(Color online) A representation of graphical solution to BA equation. The $\cot{\pi k}$ term in Eq.\ref{BA eqs} diverges at every integer $k$ and there is a root between every consecutive integer independently from $\lambda(\beta)$. By changing the value of $\lambda$, all roots can be adjusted so that the total angular momentum constraint is satisfied. } \label{cots}
\end{figure}

The $\cot{\pi k}$ term in Eq.\ref{BA eqs} diverges at every integer k, thus, regardless of the value of $\beta\:(\text{or }\lambda)$ there is a root between every consecutive integer (Fig.\ref{cots}). By changing the value of $\lambda$, all roots can be adjusted so that the total angular momentum constraint is satisfied. All the eigenstates in this problem can be labeled identically by choosing N distinct integers corresponding to the different branches of $\cot{\pi k}$ and the total angular momentum $n\in\mathbb{Z}$. The energy of an eigenstate is simply the sum of squares of all wavenumbers
\begin{equation}
E=\sum_{i=1}^Nk_i^2.
\end{equation}

For the simplest case of $\beta=0$, the ground state corresponds to $\lambda=0$ and the total angular momentum $n=0$. The roots $k$ are distributed symmetrically around zero for even N, hence, automatically satisfy the total angular momentum condition. The wavevectors for the ground state are in the N branches of \textit{cot} from $-N/2$ to $N/2$. Excitations above this ground state can be generated by two procedures. First, by changing $\lambda$, N roots which are on the same branches of \textit{cot} can be generated so as to create an eigenfunction with non-zero total angular momentum ($n\neq0$). Second, at least one of the roots can be chosen to reside on a branch that is not occupied for the ground state. For such a particle-hole excitation $\lambda$ must be adjusted to ensure the total angular momentum constraint.

Inclusion of the magnetic field affects only the total angular momentum constraint. As that constraint is defined only up to an integer (\textit{n}), the problems with values of $\beta$ differing by an integer are identical. Eigenstates for flux $\beta$ which have total angular momentum \textit{n} are also eigenstates for flux $\beta+1$ which have total angular momentum $n+1$. This is a restatement of flux quantization. We can analyze the system by considering flux values between $-1/2<\beta<1/2$.

However, in an experimental setting slowly increasing the value of the flux through the ring is a useful method to access excited states. As the flux is increased adiabatically from zero to one, the ground state evolves to an eigenstate which has its roots exactly in the same branches as the ground state, but, has a total angular momentum of minus one at zero flux (Fig.\ref{E vs beta}). The crossings between different eigenstates do not pose a problem for adiabatic evolution as only states with different total angular momentum \textit{n} can be degenerate in energy.

The BA equation Eq.\ref{BA eqs} can be very efficiently solved once the regions for the roots are determined. We used the Newton-Raphson algorithm to find a solution within a particular region. As all the roots depend monotonically on $\lambda$, another Newton-Raphson search is employed to satisfy the total angular momentum condition. We have found numerical solutions for systems of up to 10000 particles with high accuracy.

Although numerically solving the BA equation is efficient and accurate, an analytic solution can provide more insight about the physics of the system. Analytic formulae for energy, angular momentum and effective mass also would be desirable to make correspondence with experimental observations.

In the limit of strong interactions $1/c\ll1$ and large particle number $N/c\gg1$, such an analytic form can be obtained by approximating the roots of the BA equation. In this limit, because the \textit{cot} diverges quickly near integers, most of the roots are close to integers. Apart from the few roots near $k\sim\lambda$, the deviation of the root $\Delta$ from an integer $s$ is small \cite{McGuire}. Solving for this small deviation we find that the roots occur at
\begin{eqnarray}\label{k guess}
k_s^+&\!\!\!=\!&s+\frac{1}{\pi}\text{acot}\frac{2}{c}(s-\lambda),\nonumber\\
k_s^-&\!\!=\!\!&-s-\frac{1}{\pi}\text{acot}\frac{2}{c}(s+\lambda),\quad s=0,1,\ldots,\frac{N}{2}-1,\quad
\end{eqnarray}
where $acot$ is defined in the continuous region ($0,\pi$) for Eqs.\ref{k guess} to be accurate guesses. Here we have restricted $s$ to analyze the ground state and excited states with roots on the same \textit{cot} branches. Applying the total angular momentum condition we get,
\begin{eqnarray}
n-\beta&\!=\!\!\!&\sum_{s=0}^{N/2-1}k_s\nonumber\\
&\!\!\!\!=\!\!\!&\frac{1}{\pi}\sum_s^{N/2-1} \Big(\text{acot}\frac{2}{c}(s-\lambda)-\text{acot}\frac{2}{c}(s+\lambda) \Big)\nonumber\\
&\!\!\!\!=\!\!\!&\frac{c}{2\pi}\int_0^{x_F} \ud x\Big(\text{acot}(x-b)-\text{acot}(x+b)\Big),  \label{mom.rel.sum}
\end{eqnarray}
with $b=2\lambda/c$ and $x_F=(N-1/2)/c$. Here the initial assumption of strong interactions and large particle numbers allow us to approximate the sum by an integral. For the ground state and the first few excited states $n-\beta$ is small compared to $N$ and the integral can be approximated as
\begin{equation}
n-\beta=\frac{c}{2\pi}\int_{x_F-b}^{x_F+b} \ud x \text{ atan}{x}\approx\frac{cb}{\pi} \text{atan}{x_F}.\label{mom.rel.}
\end{equation}
Through this relation $b$, hence $\lambda$, is obtained for any flux value, allowing us to find expressions for all the roots in a self consistent way.

\subsection{Energy}
Using these expressions for the roots, the total energy is
\begin{eqnarray}
E&\!=\!\!\!&\sum_{s=0}^{N/2-1}k_s^2\nonumber\\
&\!\!\!\!=\!\!\!&\sum_s^{N/2-1} \bigg\{ 2s^2+ \frac{2s}{\pi}\Big(\text{acot}\frac{2}{c}(s-\lambda)+\text{acot}\frac{2}{c}(s+\lambda)\Big) \nonumber\\ &&+\frac{1}{\pi^2}\Big((\text{acot}\frac{2}{c}(s-\lambda))^2+(\text{acot}\frac{2}{c}(s+\lambda))^2\Big) \bigg\}.
\end{eqnarray}
The first term above is the total ground state energy of $N-1$ non-interacting fermions. Interactions result in the second and third terms which are first and second order corrections in our expansion. When $\Delta$'s are small, the third term is negligible. In this limit, the energy shift due to interactions is
\begin{widetext}
for repulsive interactions:
\begin{equation}
\Delta E=cb(n-\beta)+\frac{c^2x_F^2}{4}-\frac{c^2}{4\pi}\bigg\{\Big((x_F+b)^2+1\Big)\text{atan}{(x_F+b)} +\Big((x_F-b)^2+1\Big)\text{atan}{(x_F-b)}-2x_F\bigg\}, \label{deltaE analy}
\end{equation}
\begin{equation*}
\text{with } \:b=\frac{\pi(n-\beta)}{c\text{ atan}(x_F)}.
\end{equation*}
\end{widetext}

This approximate form for energy successfully reproduces numeric results for particle numbers as small as 4 throughout all the interaction range. Ground state energy as a function of interaction strength is plotted for a typical case in Fig.\ref{E vs c} for 1000 particles at $\beta=0.2$ flux. The deviation between numerical and analytical results are too small to observe in this plot.

For attractive interactions, the $\delta$-function interaction supports one bound state in one dimension. Corresponding imaginary wavevectors appear as solutions of the BA equation. For $k=\alpha+i\sigma$ with $(\alpha,\sigma)\in\mathbb{R}$, the BA equation has only two roots with $\sigma\neq0$. The charged particle is bound with only one of the background fermions. When $1/|c|\ll1$, the complex roots are at $k=\lambda\pm i c/2$ while the rest of the roots preserve their form of Eqs.\ref{k guess}. If the bound state is narrow, Pauli repulsion between the fermion in the bound pair and background fermions decreases the effective interaction. Within these approximations, we analytically calculate the total energy,
\begin{widetext}
for attractive interactions:
\begin{equation}
\Delta E=-\frac{c^2(b^2+1)}{2}+cb(n-\beta)+\frac{c^2x_F^2}{4}+\frac{c^2}{4\pi}\bigg\{\Big((x_F+b)^2+1\Big)\text{atan}{(x_F+b)} +\Big((x_F-b)^2+1\Big)\text{atan}{(x_F-b)}-2x_F\bigg\}, \label{deltaE analy -c}
\end{equation}
\begin{flushright}
\begin{equation*}
\text{with } \:b=\frac{(n-\beta)}{c\Big(1-\frac{1}{\pi}\text{atan}(x_F)\Big)}.
\end{equation*}
\end{flushright}
\end{widetext}

\subsection{Angular Momentum}
To understand the physics of the system and make correspondence to possible experiments, it is important to calculate other measurable quantities. In particular, for this system we are interested in how the dynamics of the impurity particle is affected by the fermion background. To this end, it is instructive to calculate angular momentum carried by the impurity $L_1$ and the related effective mass. As the impurity is interacting with the fermions, this effective mass is not only the mass of the impurity but also gets a contribution from the fermions dragged along with it. Such a compound object is generally called the polaron state or dimer state especially for attractive interactions.

As stated above, one of the most interesting physical quantities in this system is the angular momentum carried by the charged particle, represented by the operator $\hat{P}_1=-i\frac{\partial}{\partial x_1}$. As this particle is coupled to the external magnetic field, $\hat{P}_1$ is the canonical momentum not the kinetic momentum. However, canonical momentum is the quantity that is generally measured by expansion imaging in artificial magnetic field experiments. The expectation value of $\langle\hat{P}_1\rangle=L_1$ is easily obtained by taking the derivative of the total energy with respect to flux,
\begin{equation}
L_1=\frac{-1\:}{2}\,\frac{\partial\Delta E}{\partial\beta}.
\end{equation}

\begin{figure}
\includegraphics[width=0.48\textwidth]{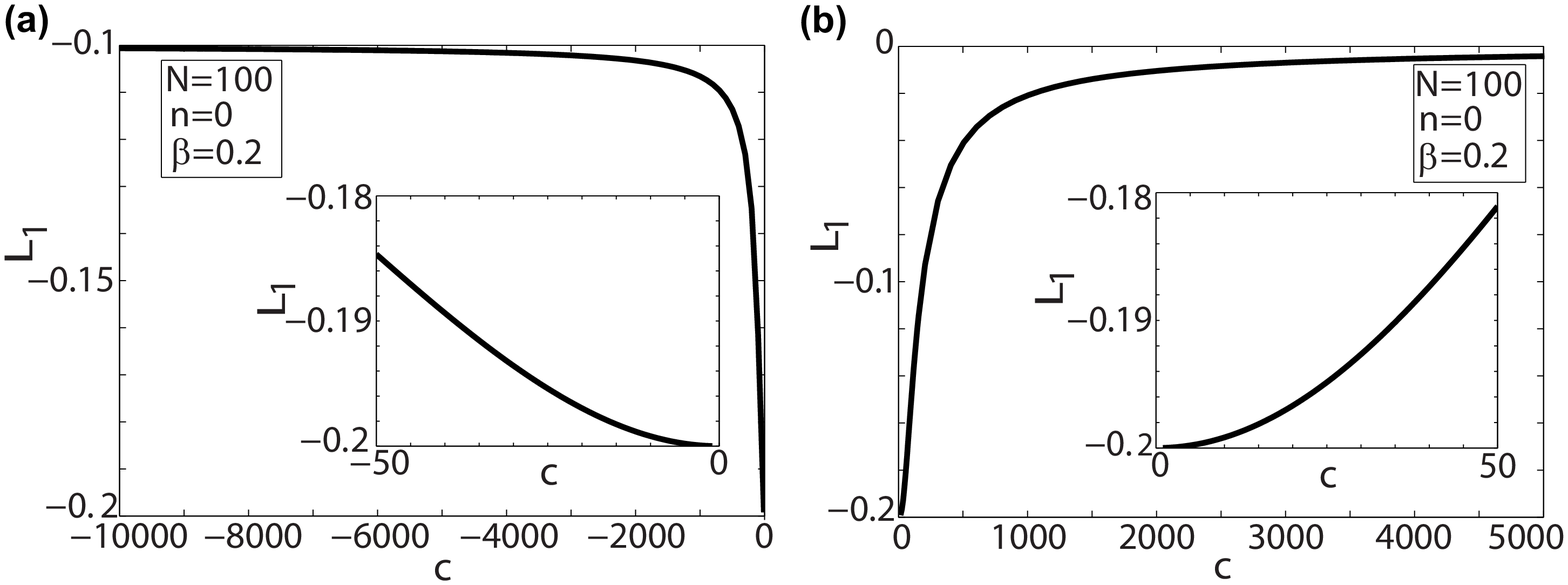}
\caption{ Angular momentum of the impurity vs. interaction strength for $N=100$ particles for (a) attractive, (b) repulsive interactions from the analytic calculation, numerical solutions produce the same results. In the non-interacting limit, the impurity carries all the angular momentum ($n-\beta$). $L_1$ saturates to almost zero for infinitely strong repulsive interactions as the total angular momentum is shared equally between all particles. The same behavior holds for excited states. For strongly attractive interactions, $L_1$ saturates to half the total value signifying dimer formation with one background particle. The insets in both figures focus on the weak interaction limit. } \label{L1 vs c1}
\end{figure}
Using the approximate form for the energy Eq.\ref{deltaE analy}, we obtain
\begin{widetext}
\begin{equation}\label{L1eq}
L_1=\frac{\pi(n-\beta)}{\text{atan}{x_F}}-\frac{c}{4\text{atan}{x_F}}\bigg\{(x_F+b)\text{atan}{(x_F+b)} -(x_F-b)\text{atan}{(x_F-b)} \bigg\}.
\end{equation}
\end{widetext}
This form is valid for positive $c$ and easy to interpret. In the non-interacting limit, the canonical momentum of the charged particle is fixed by the external flux. Hence, all the angular momentum is carried by the charged particle. As interaction is turned on, the charged particle drags the background fermions and transfers some of its angular momentum to them. Stronger interactions increase the fraction of the transferred angular momentum and in the limit of infinitely repulsive interactions angular momentum is equally shared by N particles. On the other hand, for strong attractive interactions, $L_1$ saturates to half of the angular momentum in the system proving the formation of a dimer with one background fermion.
\begin{figure}
\includegraphics[width=0.45\textwidth]{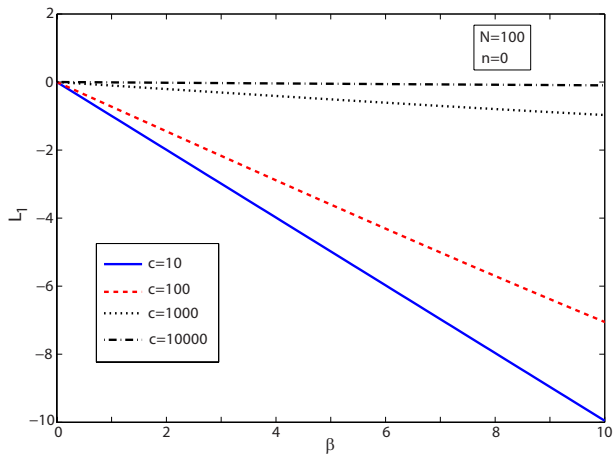}
\caption{(Color online) The angular momentum of the charged particle vs. flux at varying interaction strength for $N=100$ particles. As expected $L_1$ depends linearly on flux. The slope of the line decreases with increasing interaction strength indicating higher values of effective mass of the impurity.} \label{L1 vs beta}
\end{figure}

The behavior of $L_1$ is displayed in Fig.\ref{L1 vs c1} and Fig.\ref{L1 vs beta} as a function of $c$ and $\beta$. Even for $N=100$ particles, the difference between numerical calculation of the derivative and the expression given above is negligible. As a function of interaction strength, the rapid decrease and eventual saturation of $L_1$ validates the scenario discussed above. The linear dependence on flux is expected, however, the slope of $L_1$ decreases as interaction gets stronger. This slope carries valuable information as it is related to the effective mass of the composite excitation formed by the impurity and background fermions.

\subsection{Effective Mass}
\begin{figure}
\includegraphics[width=0.48\textwidth]{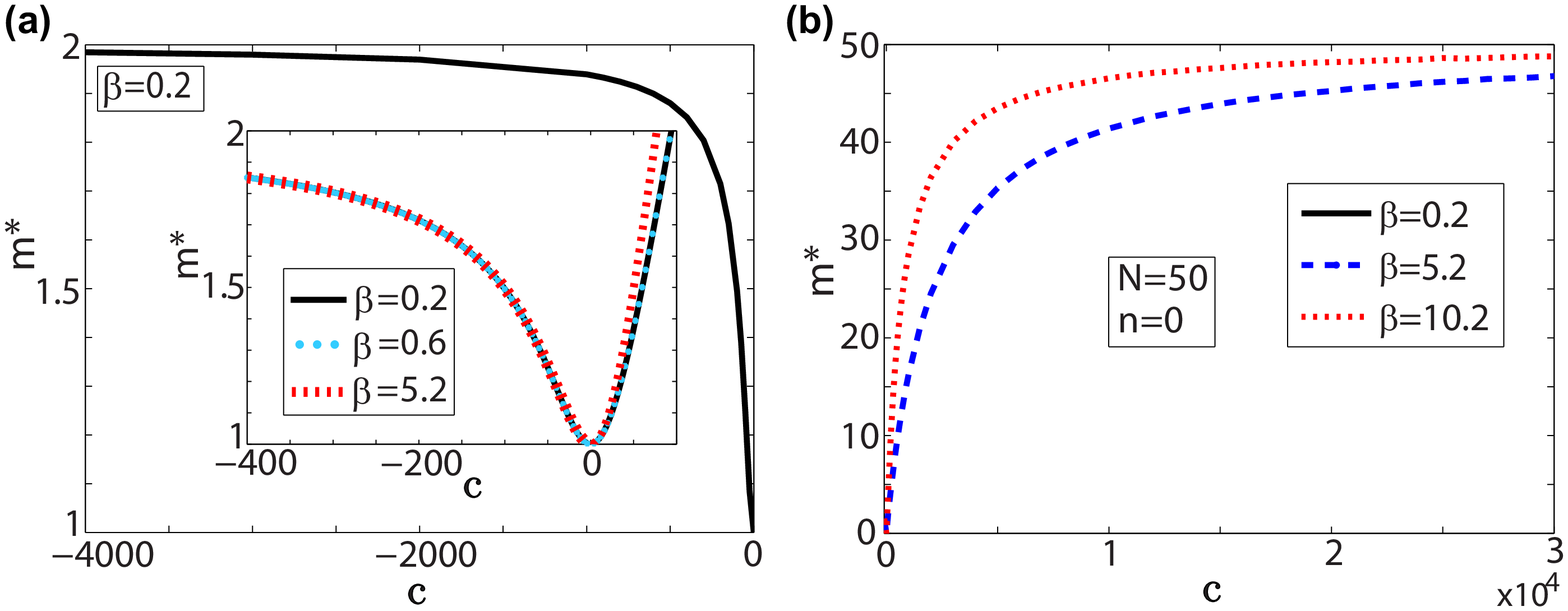}
\caption{ (Color online) Effective mass of the impurity vs. interaction strength for $N=50$ particles and zero total angular momentum $n=0$. (a) For attractive interactions, $m^*$ saturates to twice the mass of the impurity due to the formation of a tightly bound pair. The inset shows the behavior around zero interaction in more detail. For attractive interaction, the effective mass (given by Eq.\ref{ddE}) is almost insensitive to flux change. (b) For repulsive interactions, $m^*$ converges to $N$. As flux $\beta$ increases, this saturation gets faster. The dependence on the flux is more prominent for small particle numbers.   } \label{m* vs c}
\end{figure}
We define the effective mass as
\begin{equation} \label{m* eq}
m^*=\frac{2}{\frac{\partial^2\Delta E}{\partial\beta^2}}.
\end{equation}
In the non-interacting limit, the effective mass is equal to $m$, however, its behavior is very different for attractive and repulsive interactions. As repulsive interactions are increased, it gets harder for the impurity to tunnel through the fermions and the dragged particles increase the effective mass. The increase in the effective mass saturates only when all the particles are moving together with the impurity. Thus, at large repulsive interaction, the effective mass reaches $Nm$. For weak attractive interactions, the first effect is once again the drag increasing the effective mass. However, the attractive $\delta$-function has a single bound state in one dimension. Thus, the impurity captures one of the background fermions and as the size of the bound state gets smaller, Pauli exclusion effectively repels the other fermions. The effective mass for attractive interactions increases and reaches $2m$ for infinitely attractive interaction where a dimer is formed from the impurity and one fermion. For attractive interactions, the analytical expression in the strongly interacting limit is useful to calculate the dimer mass,
\begin{widetext}
\begin{eqnarray} \label{ddE}
\frac{\partial^2\Delta E}{\partial\beta^2}&=&\frac{2}{1-\frac{\text{atan}{(x_F)}}{\pi}}-\frac{1}{(1-\frac{\text{atan}{(x_F)}}{\pi})^2}\nonumber\\ &&+\frac{1}{2\pi(1-\frac{\text{atan}{(x_F)}}{\pi})^2} \bigg\{\text{atan}{(x_F+b)}+\text{atan}{(x_F-b)}+\frac{x_F+b}{1+(x_F+b)^2}+\frac{x_F-b}{1+(x_F-b)^2} \bigg\}.
\end{eqnarray}
\end{widetext}
We calculated the effective mass numerically and analytically. For the ground state, the above scenario is validated by these calculations (see Fig.\ref{m* vs c}). The dependence of the effective mass on the external magnetic field is strongest for small particle number as this limit is the strongly interacting limit in one dimension. As the number of fermions increases, effective mass in the ground state has weak dependence on $\beta$. In this case, effective mass is essentially determined locally as the ability for the impurity particle to complete a full rotation is hampered.

\begin{figure}
\includegraphics[width=0.45\textwidth]{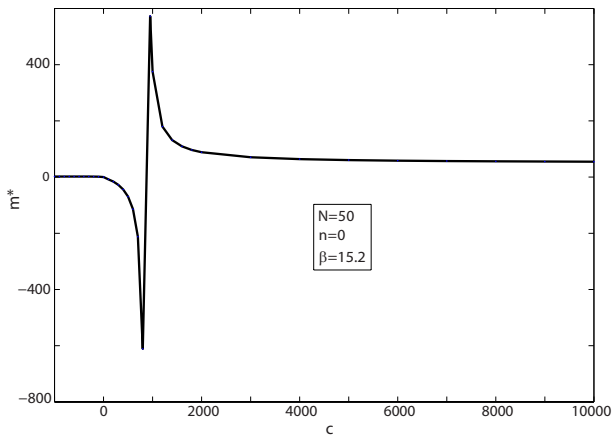}
\caption{ Resonant behavior in $m^*$ for $\beta$ comparable to $N$. When the drag effect applied by the background particles overcomes the driving force of the magnetic field, $m^*$ can become negative. When the second derivative of the energy with respect to flux becomes zero, $m^*$ diverges. This divergence does not change the infinitely strong interaction limit. } \label{m* vs c resonance}
\end{figure}
The utility of an external magnetic field is the access it provides to excited states through adiabatic pumping. Excited states in this system are expected to be stable due to angular momentum conservation. It is thus reasonable to expect effective mass measurements to be carried out on such states in a cold atom setting. For the excited states, with angular momentum $|n|<N/4$ the main effect is faster saturation of effective mass as $c$ increases. However, for higher excited states, there is resonant behavior (Fig.\ref{m* vs c resonance}). Due to the nature of BA solution, a state for which all the roots are on the \textit{cot} branches from $-N/2$ to $N/2$ can have at most $N/2$ units of angular momentum. When the total angular momentum of an excited state is comparable to particle number N, the sharing of this angular momentum between the impurity and the background is limited by this constraint. Thus, it is possible for this system to support negative effective mass if the external force acted by the magnetic field is overcome by the back reaction from the fermions. We numerically find this behavior for both low and high particle number (see Fig.\ref{m* vs c resonance}). Experimentally this effect should be more accessible for small number of particles as it is easier to pump angular momentum comparable to particle number.
\begin{figure}
\includegraphics[width=0.45\textwidth]{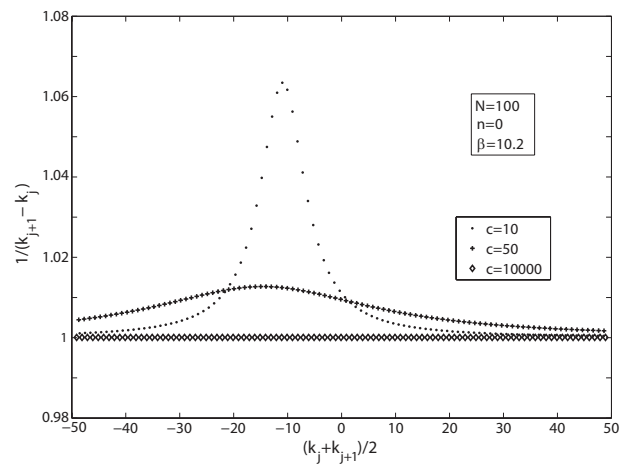}
\caption{ Effective momentum density in k-space for different interaction strengths. In the strongly interacting limit, the distance between adjacent BA roots (wavevectors) is one. For small c, the roots are closer to each other around $(n-\beta)$. The impurity carrying $n-\beta$ units of angular momentum in the non-interacting limit first disturbs the fermions which are momentum matched to that value. } \label{k-dist beta10.2}
\end{figure}

\subsection{Correlations}
Apart from the single particle properties related to the impurity, it is instructive to look at global properties to understand how the external particle disturbs the one dimensional Fermi liquid. A common way to visualize the disturbance in the Fermi sea is to plot the deviation of the distance between the BA roots (wavevectors) from one. For an undisturbed Fermi sea, this deviation is always one. For a weakly interacting impurity, the deviation is confined to a narrow region in k-space around $n-\beta$ (see Fig.\ref{k-dist beta10.2}). This is expected as the impurity carrying $n-\beta$ units of angular momentum in the non-interacting limit first starts dragging fermions which are matched in momentum. As the strength of repulsion increases, so does the effected region in k-space, however, the deviation gets smaller. For infinitely repulsive interactions, the impurity becomes indistinguishable from the background fermions (Fig.\ref{k-dist beta10.2}). For highly excited states where $n$ is comparable to N, particle-hole excitations complicate this picture similar to the effect we discussed for the effective mass.

Another important physical property is the two-particle correlation function. Although for $\delta$-function interactions only the value of this function at zero determines the interaction energy, its general form is experimentally accessible through Hanbury-Brown-Twiss\cite{HanburyBrownTwiss} type measurements. This correlation also can be regarded as the real-space form of the bound state created by the impurity. To calculate the two-particle correlation function, we need to determine the coefficients of the plane waves in each region. Following McGuire \cite{McGuire} we choose the first coefficient in the first region $x_1<x_2<\ldots<x_N$,
\begin{equation}
(123..N)_{123..N}=(1-e^{i2\pi k_1}).
\end{equation}
Other coefficients in this region determined by BCs yield very similar expressions. The wavenumber associated with the distinguishable particle appears in the exponent and the sign of the permutation multiplies the coefficient:
\begin{eqnarray}
(213..N)_{123..N}&=&-(1-e^{i2\pi k_2}),\\
(312..N)_{123..N}&=&(1-e^{i2\pi k_3})\\
&\vdots& \nonumber
\end{eqnarray}
The coefficients in other regions are related to the same coefficient in the first region with a phase factor determined by the PBCs. This phase factor is a full circle rotation around the ring of the particles that $x_1$ has to pass to be in the given region. Thus, the momenta belonging to the particles that $x_1$ has passed multiply the coefficient, e.g.
\begin{equation}
(21354..N)_{2314..N}=(1-e^{i2\pi k_2})e^{i2\pi(k_1+k_3)}\ldots
\end{equation}
In the simple form, the wave functions are not normalized, but we normalize the correlation function at the end. The two-particle correlation function in any state is given as
\begin{equation}
g_{12}(x_1,x_2)=\int_0^{2\pi}\ud x_3\cdots\int_0^{2\pi}\ud x_N \Psi^*\Psi.
\end{equation}

\begin{figure}
\includegraphics[width=0.47\textwidth]{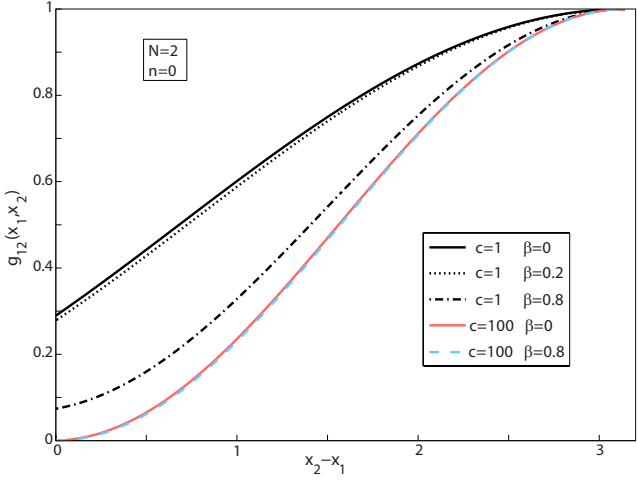}
\caption{(Color online) Two-particle correlation function for $N=2$ particles. As expected, $g_{12}$ at zero separation decreases with increasing interaction strength. For weak interactions, the correlation function at zero decreases with increasing flux. However, for strong interactions, the correlation is almost insensitive to flux change due to the fermionization of the charged particle. } \label{g12N2 fig}
\end{figure}

For the $N=2$ particle case, the correlation function is simply the absolute square of the wave function. As could be expected, the correlation function is highly affected by the flux for the two-particle case. Using the numerically and analytically found wavenumbers in the expression,
\begin{align}
g_{12}(x_1,x_2)&=4-2\cos(2\pi k_1)-2\cos(2\pi k_2)\nonumber\\
&-8\sin(\pi k_1)\sin(\pi k_2)\cos(k_1-k_2)(x_2-x_1-\pi),
\end{align}
we observe that inclusion of the flux generally decreases the two-particle correlation function (Fig.\ref{g12N2 fig}). However, if the interactions are strong enough so that the particles are almost fermionized, this decrease is very small. It is also notable that although the flux breaks time-reversal symmetry and the wave functions choose a direction on the ring, correlation function is even with respect to $x_1-x_2$. This property holds for any particle number.

\begin{figure}
\includegraphics[width=0.47\textwidth]{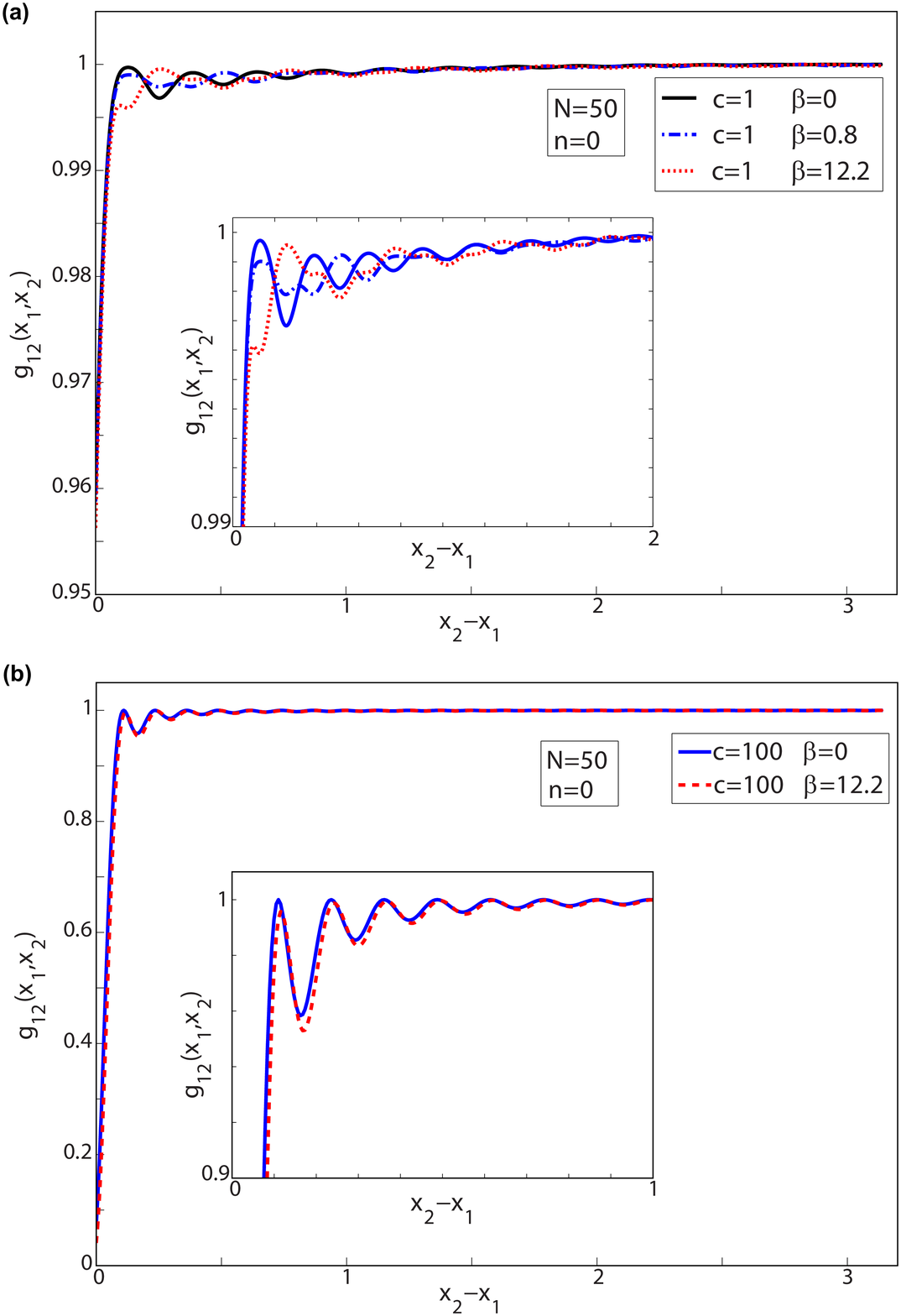}
\caption{(Color online) Two-particle correlation function for $N=50$ particles. (a) For weak interactions, Friedel oscillations occur as interference of two waves with wavelengths related to $k_F-\beta$ and $k_F+\beta$. (b) At strong interactions, the correlation becomes zero at zero separation since the impurity is effectively indistinguishable and $g_{12}$ and the frequency of Friedel oscillations are almost insensitive to flux change. } \label{g12N50 fig}
\end{figure}
For the general N particle case, we arrange the wave function in a better form to evaluate the integrals. We assign two wavenumbers to $x_1$ and $x_2$, and the rest of the particles are represented by a Slater determinant since they are indistinguishable fermions. For example, if we have $k_1,k_4$ associated with $x_1,x_2$ respectively, the Slater determinant is represented by $\mathbb{D}_{14}$ indicating the use of all wavenumbers except $k_1$ and $k_4$ in the exponents,
\begin{equation}
\mathbb{D}_{14}=
\begin{vmatrix}
e^{ik_2x_3} & e^{ik_3x_3} & e^{ik_5x_3}\,\ldots \\\\
e^{ik_2x_4} & e^{ik_3x_4} & e^{ik_5x_4}\,\ldots \\\\
\vdots & & & \\\\
e^{ik_2x_N} & e^{ik_3x_N} & e^{ik_5x_N}\,\ldots
\end{vmatrix}.
\end{equation}
Hence, the wave function in the first region can be written as,
\begin{align}
\Psi=&\Big((12\ldots N)_{12\ldots N}+(21\ldots N)_{12\ldots N} \Big)\mathbb{D}_{12}\nonumber\\
&\:+\Big((13\ldots N)_{12\ldots N}+(31\ldots N)_{12\ldots N} \Big)\mathbb{D}_{13}+\ldots
\end{align}
Integrating $\Psi^*\Psi$ over $x_3,\ldots,x_N$, the Slater determinants are orthogonal in large particle number limit, as the outer roots of \textit{cot}'s are very close to integer values. The correlation function is then expressed as a sum over pairs of momenta associated with $x_1$ and $x_2$,
\begin{align}
g_{12}(x)=&\sum^N_{t<s}\sum^N|(ts\ldots N)|^2+|(st\ldots N)|^2\\
&+2Re\Big\{(ts\ldots N)^*(st\ldots N)e^{i(k_t-k_s)x} \Big\},
\end{align}
where $x=x_2-x_1$. Here, $x>0$ but the symmetry of the correlation function for $x<0$ can be easily seen by using the region $x_2<x_3\ldots x_N<x_1$ instead of the first region. Finally, the correlation function is normalized to average density on the ring so that it saturates to one. The correlation function for $N=2$ and $N=50$ particles are given in Fig.\ref{g12N2 fig} and Fig.\ref{g12N50 fig} respectively, for different interaction strengths and flux. As expected, the correlation function at zero separation $g_{12}(0)$ decreases with increasing interaction strength until it saturates to zero. The other important feature of the correlation function is the Friedel oscillations \cite{Friedel} reflecting the sharpness of the Fermi surface in one dimension.

The two-particle correlation function is a local quantity while the external flux changes the system properties globally. For any pair to feel the effect of the flux, one of the particles must go a full circle through the ring. Hence, as could be expected, the effect of the artificial magnetic field on the correlation function decreases as the number of particles increases or if they interact strongly. Even for the lowest lying excited states, we find the primary effect of the flux is on the Friedel oscillations while the shape of the correlation hole is unchanged.

\begin{figure}
\includegraphics[width=0.48\textwidth]{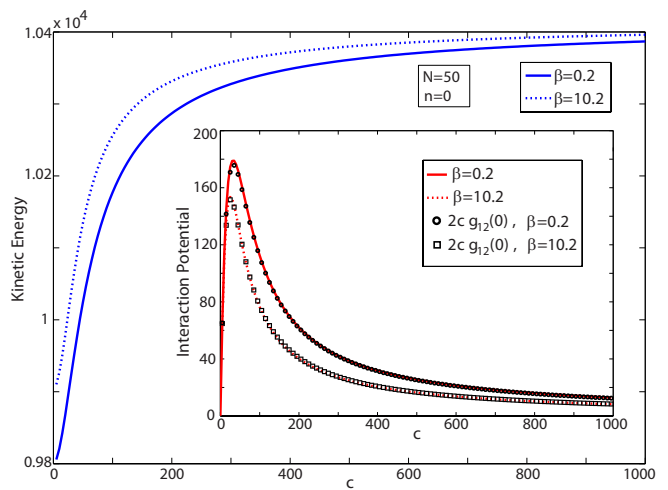}
\caption{ (Color online) Kinetic energy of the particles vs. interaction strength for $\beta=0.2$ and $\beta=10.2$ for $N=50$ particles. The inset shows the interaction potential contribution to the total energy. The initial increase in interaction energy follows the increase in the interaction strength. However, beyond a certain strength, the tendency of the fermions to avoid the impurity is more dominant. These plots are obtained by taking the derivative of the total energy with respect to $c$. Alternatively, the interaction potential energy is also obtained by using the two-particle correlation function at zero separation. Both results are plotted in the inset showing remarkable agreement. }  \label{Kin Pot}
\end{figure}

Finally, we calculate a thermodynamic quantity which is also related to $g_{12}(0)$. Derivative of the energy with respect to interaction strength $c$ gives us interaction potential, so, the kinetic and interaction contributions to the total energy can be separated. In Fig.\ref{Kin Pot}, one can notice that the interaction energy makes a peak and then decreases for increasing $c$. The initial increase is expected, however, as interactions become stronger, the tendency of the fermions to avoid the impurity dominates and the impurity is effectively fermionized. This is apparent in the $\delta$-function BC Eq.\ref{discont genel}. Additionally, the interaction potential is equal to the correlation function at zero $g_{12}((x_2-x_1)=0)$ times the interaction strength which reproduces the results obtained by taking the derivative of the total energy.

\section{CONCLUSION}
The problem of a single impurity interacting with a fermion background has attracted attention of the condensed matter society for years. In this paper, we argue that an artificial gauge field coupling exclusively to the impurity is an effective tool to probe the physics of this system at any interaction strength. We consider a Fermi gas in a narrow ring trap and an artificial magnetic field coupling only to a single impurity. We solve this system exactly by using the BA for contact interactions and calculate the dependence of measurable quantities on the external magnetic flux. We observe this dependence for total energy, angular momentum of the charged particle, effective mass and the two-particle correlation function.

Using an artificial magnetic field in this system has two advantages. The usual measurement tools such as expansion imaging become probes of thermodynamic quantities by comparing measurements at different flux values. For example, the change of the momentum carried by the impurity caused by the magnetic field is a direct probe of the effective mass of the impurity. The second advantage is obtained by adiabatically increasing the flux value. Although the Hamiltonian of the system is periodic with flux, adiabatic evolution connects the ground state at zero flux to excited states at integer flux. In a cold atom experiment, such excited states can be expected to have long lifetimes due to total angular momentum conservation. Thus, we have calculated the physical properties for not only the ground state but also for excited states adiabatically connected to it.

Our results show that the system can be described by a simple physical picture. The charged particle interacting with the background particles drags them along with itself around the ring. In the non-interacting limit, all of the angular momentum in the system is carried by the impurity. As interactions are turned on, fermions which are close to the impurity in momentum are disturbed more and start to gain momentum. At the limit of infinitely repulsive interactions, the charged particle is effectively indistinguishable from the background and the total momentum is shared equally between all particles. The pair correlation function also confirms this picture. The value of the correlation function near zero is mostly insensitive to the external flux while away from the correlation hole frequency of the Friedel oscillations sensitively depends on it. For strongly repulsive interactions, the effective mass saturates the total mass of the particles since it is dragging all of the background fermions along with itself around the ring.

For attractive interactions, the impurity forms a bound pair with one of the fermions. The effect of dimer formation can be clearly seen in the angular momentum and the effective mass. For infinitely strong attractive interactions, angular momentum carried by the impurity saturates half the value of total angular momentum and the effective mass saturates twice the mass of the particle which confirm the presence of the dimer as a composite particle.

The physical properties calculated in this paper are experimentally accessible through the standard tools of ultracold atom experiments. While artificial magnetic fields have been demonstrated in a variety of settings, they have not been used in combination with a toroidal trap to our knowledge. We believe our exact results would be relevant for such an experiment.

\begin{acknowledgments}

F.N.\"{U}. is supported by T\"{u}rkiye Bilimsel ve Teknolojik Ara\d{s}t{\i}rma Kurumu (T\"{U}B\.{I}TAK) Scholarship No. 2211. M.\"{O}.O. was supported by T\"{u}rkiye Bilimsel ve Teknolojik Ara\d{s}t{\i}rma Kurumu (T\"{U}B\.{I}TAK) Grant No. 112T974. B.H. is supported by T\"{u}rkiye Bilimsel ve Teknolojik Ara\d{s}t{\i}rma Kurumu (T\"{U}B\.{I}TAK) Grant No. 113F334.

\end{acknowledgments}

\bibliography{PaperPRAResubmit}

\end{document}